# High-density and scalable graphene Hall sensor arrays through monolithic CMOS integration

Vasant Iyer[1,2,*†], Nishal Shah[3], A.T. Charlie Johnson[4], David A. Issadore[1,3*], Firooz Aflatouni[1*]

[1] Department of Electrical and Systems Engineering, School of Engineering and Applied Science, University of Pennsylvania, Philadelphia, Pennsylvania, United States

[2] Querrey Simpson Institute for Bioelectronics, Northwestern University, Evanston, Illinois, United States

[3] Department of Bioengineering, School of Engineering and Applied Science, University of Pennsylvania, Philadelphia, Pennsylvania, United States

[4] Department of Physics and Astronomy, School of Arts and Sciences, University of Pennsylvania, Philadelphia, Pennsylvania, United States

*Co-corresponding author: vasant@northwestern.edu

*Co-corresponding author: issadore@seas.upenn.edu

*Co-corresponding author: firooz@seas.upenn.edu

†Lead contact

## ABSTRACT

Electronic devices made from two-dimensional materials (2DMs) significantly outperform their silicon counterparts; however, silicon CMOS technology remains commercially predominant as it offers the capability to operate dense arrays of devices in a scalable fashion. In particular, graphene Hall sensors (GHSs) offer great improvements in magnetic field sensitivity and resolution compared to silicon Hall-effect sensors, making them extremely appealing for magnetic field imaging and biosensing. At present, GHS arrays have limited scalability compared to silicon CMOS since they require planar routing for biasing and multiplexing. In this work, we explore strategies to realize high-density graphene Hall sensor arrays by vertically connecting GHSs with silicon CMOS biasing and multiplexing circuitry, allowing the routing and circuitry to scale with the array. We investigate the importance of design choices in the chip layout and post-fabrication process in maximizing the reliability of graphene transfer onto mm-scale CMOS dies. Our experimental results validate the success of the integration process by showing for the first time that GHSs can be monolithically integrated with CMOS with high yield to form sensor arrays. We expect that these results will lead to further improvements in magnetic sensing technology and broader advancements in large-scale heterogeneous 2DM-CMOS systems.

# INTRODUCTION

Two-dimensional materials (2DMs) such as graphene have garnered extraordinary interest from the device community in the past two decades due to their exceptional electrical and optical properties.[1,2] Recently reported 2DM devices such as transistors,[3,4] memory elements,[5] biological and chemical transducers,[6,7] photodetectors,[8] and modulators[9] offer better performance and new functionalities compared to state-of-the-art semiconductor devices. A key goal for commercializing 2DM devices is to incorporate them into larger-scale computational and sensing systems that benefit from their special properties.[10,11] While advancements in synthesis and transfer techniques have made it feasible to mass produce 2DM devices, the widespread usage of these systems remain limited by several factors. These include low device yield and performance uniformity,[12–15] as well as the need for laboratory equipment such as automated probe stations to concurrently operate large numbers of devices.

A promising approach towards system scaleup is by integrating 2DM devices with silicon complementary metal-oxide-semiconductor (CMOS) integrated circuits (ICs), which are used ubiquitously in consumer electronics. Integrating 2DMs with silicon CMOS aims to synergize the strengths of each individual material platform - the high performance and multifunctional nature of 2D materials with the speed, reliability, low unit cost, and mature circuit architectures of CMOS technology.[16] Early 2DM-CMOS systems employed hybrid integration (**Fig 1A**), in which 2D material devices are fabricated on one substrate and connected to CMOS circuits using planar routing and interconnections such as wirebonds.[17–19] Hybrid integration decouples the fabrication constraints of each material system, allowing the use of state-of-the-art 2DM synthesis techniques to maximize the performance gain from using 2DM elements; however, the area overhead imposed by planar routing and interconnection scales poorly with system size, limiting the scalability of the hybrid approach.

Compared to hybrid integration, monolithic (i.e. single-chip) 2DM-CMOS integration (**Fig 1B**) allows for vertical readout and interconnection, enabling 2DM devices to be massively parallelized and packed onto a chip with much higher densities. Although this approach involves additional fabrication complexity due to the imposition of thermal and chemical compatibility constraints, monolithic integration has become feasible within the past few years following significant advancements in CMOS-compatible synthesis and transfer techniques.[20–22] Recently demonstrated monolithic 2DM-CMOS systems have been implemented by integrating 2DM devices within the back-end-of-line (BEOL) layers of CMOS ICs manufactured using foundry processes.[23,24] The systems cover a wide range of applications and include broadband imagers,[25,26] crossbar arrays for in-memory computing,[27] and sensor arrays to detect gases and biomolecules.[28–30] The most successful of these demonstrations in terms of device yield used large CMOS dies or whole-wafer (>200 mm$^2$) processing,[25,27,30] often requiring high prototyping cost and/or customization of the foundry process. By comparison, integration efforts onto relatively inexpensive mm-scale CMOS chips procured through multi-project wafer runs have reported lower yield (~50%) and transfer reliability due to the challenges of transferring 2D materials onto small dies.[28,31–33]

In this work, we consider the opportunities of monolithic 2DM-CMOS integration for implementing magnetic sensing arrays using graphene Hall-effect sensors (GHSs). Microscale magnetic sensors employing the Hall effect are ubiquitous in automotive and navigation applications requiring high linearity and bandwidth. These sensors are predominantly made from silicon to reduce fabrication cost, but they suffer from poor field resolution and sensitivity due to silicon's low charge carrier mobility.[34,35] Hall sensors implemented with high-mobility III-V semiconductors exhibit much better performance and are better suited to precision applications, but their scalability is limited by fabrication cost.[36,37] Graphene Hall sensors possess comparable field sensitivity and resolution performance to III-V devices while remaining compatible with silicon fabrication.[38–40] Since GHSs combine the advantages of both existing Hall sensing platforms, they are an appealing choice for implementing high-performance magnetic sensing arrays with applications in scanning magnetometry,[41] mapping current distributions in batteries and other electrical devices,[42,43] and sensing magnetically labeled cells and particles in biomedical diagnostics.[44–46] As with other 2DM devices, the feasibility of using graphene Hall sensors in large arrays is limited by the need for external biasing, multiplexing, and readout circuits; additionally, performance variability and drift in GHS arrays remain significant.[47] Previous studies have attempted to address these issues by integrating graphene Hall sensors on top of CMOS circuits and interconnecting them with wirebonds.[17,48] While these demonstrations showed promising sensing results with single CMOS-integrated graphene Hall sensors, similar results could not be shown for GHS arrays due to the scalability challenges associated with planar interconnection.

In this work, we explored the use of CMOS integrated circuits to implement high-density GHS arrays using a monolithic integration approach. We designed a proof-of concept IC in commercial 180nm CMOS technology which supported 32 Hall sensing sites and integrated graphene into the BEOL layers of the chip. We found that the feasibility of graphene integration onto CMOS could be enhanced through 1) designing the chip layout and post-fabrication process to integrate graphene onto the BEOL inter-layer dielectric (ILD) and 2) modifying the conventional graphene transfer process to ensure conformal contact with the CMOS chip. The transferred graphene was analyzed using several physical and electrical characterization techniques, validating the integration process in terms of graphene quality and graphene-CMOS contact. Finally, magnetic measurements were used to assess the Hall sensing performance of the integrated devices, showing high device yield and responsive graphene Hall elements. This work shows for the first time that graphene Hall sensors can be monolithically integrated with silicon CMOS, paving the way towards large-scale magnetic sensing arrays that combine highly sensitive GHSs with CMOS-based readout and conditioning circuits.

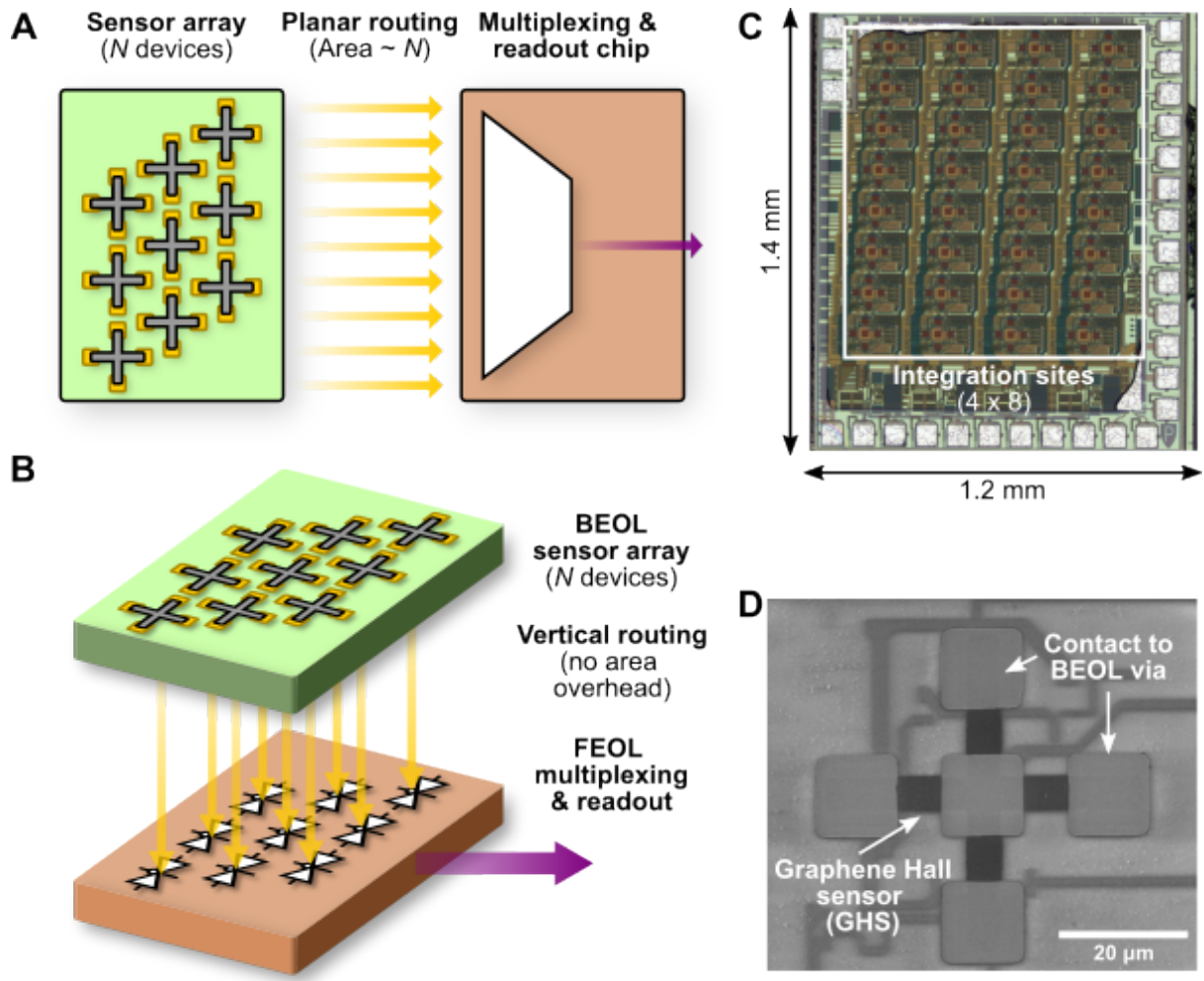

**Figure 1: Dense and scalable graphene Hall sensor arrays through monolithic CMOS integration.** A) Schematic representation of *N* graphene sensors connected to CMOS-based multiplexing and readout circuits using hybrid (multi-substrate) integration. B) Schematic representation of *N* graphene sensors monolithically integrated into the back-end-of-line (BEOL) layers of a CMOS chip whose front-end-of-line (FEOL) layers host multiplexing and readout circuits. C) Micrograph of the prototype CMOS chip designed for graphene integration (top metal fill over the sensing region removed for visualization purposes). D) Scanning electron micrograph of a single sensing site after CMOS post-processing, graphene integration, and Hall sensor patterning.

## RESULTS

### CMOS chip design for GHS integration

The CMOS-GHS chip was designed as a scalable array of sensing pixels in a 180nm standard CMOS process (Taiwan Semiconductor) (**Fig 1C**). Each pixel consisted of a GHS integrated within the BEOL layers of the CMOS chip (**Fig 1D**), along with biasing and multiplexing circuitry implemented in the underlying silicon layer. The GHS was sized to occupy 100 µm x 100 µm of chip area, including device contacts. The location and geometry of each GHS was defined during chip design by the placement of five contacts (four device contacts and one backgate contact, each 20 µm x 20 µm) connecting the GHS to the biasing and multiplexing circuitry. The prototype chip included 32 pixels (4 groups x 8 pixels/group) within a total chip area of 1.4 mm x 1.2 mm including wirebonding pads for external connections.

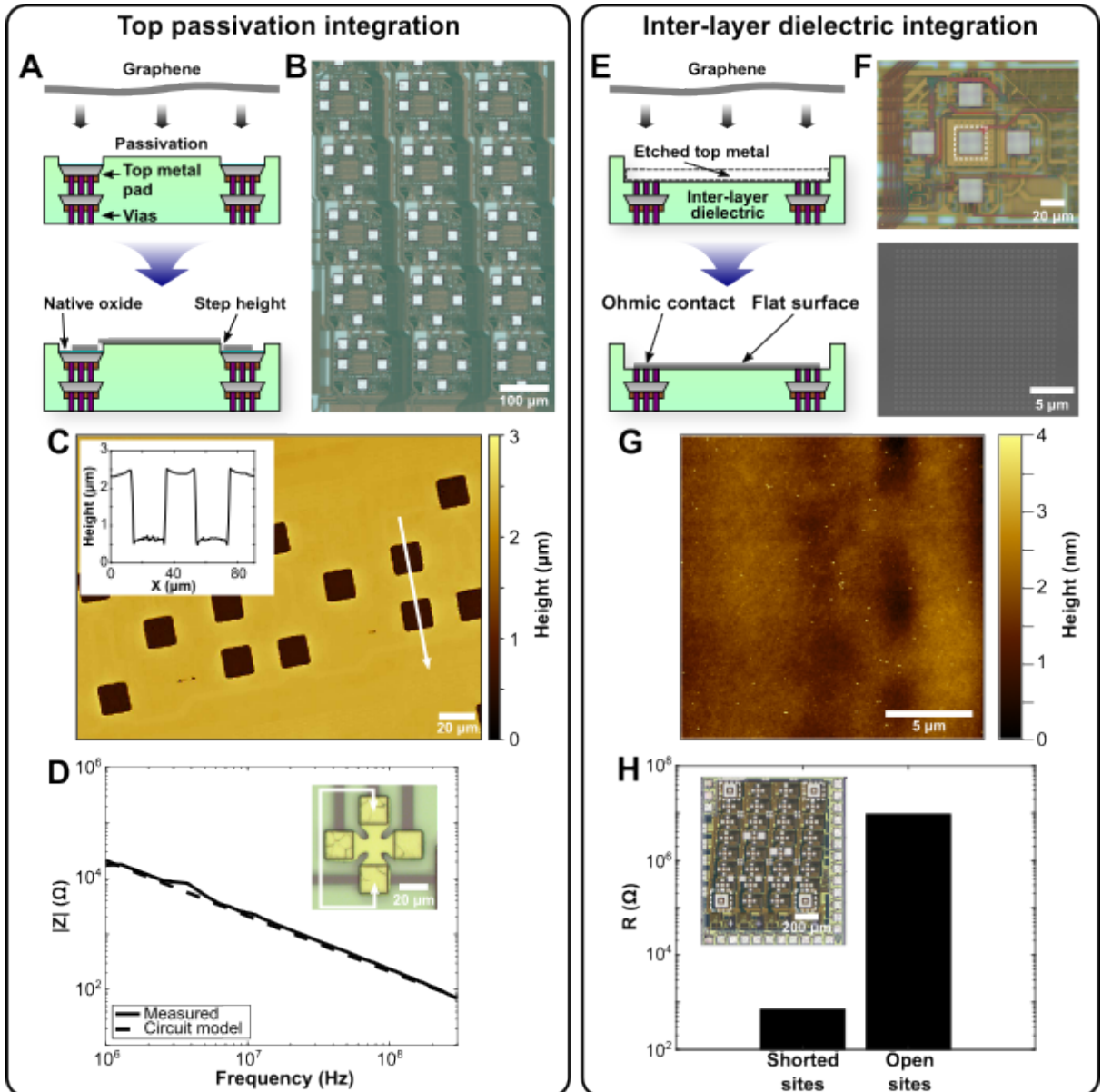

**Figure 2: Comparison between graphene-CMOS integration approaches.** A) Schematics of graphene integration onto the CMOS top passivation. B) Micrograph of a CMOS chip containing several sites designed for top-passivation integration of graphene Hall sensors. C) Topography of the chip surface measured with optical profilometry. The arrow indicates the location of the line scan shown in the inset. D) Impedance of a top-passivation sensing site shorted by an evaporated Ti/Au trace. The inset shows a micrograph of the shorted sensing site. E) Schematics of graphene integration onto the inter-layer dielectric (ILD) accessed after etching a top-metal sacrificial layer. F) Micrograph of a sensing site designed for graphene integration onto the ILD (top), shown with the top-metal fill and diffusion barrier removed. A scanning electron micrograph of the dashed box (bottom) shows the surface of the vias exposed after etching the top-metal fill. G) Atomic force microscope map of the ILD surface. H) Comparison of resistances measured at sensing sites shorted by evaporated Ti/Pd traces (inset) versus sites left open.

**Comparison of graphene-CMOS integration approaches**

A major challenge of this work was choosing a graphene-CMOS integration process that would maximize the reliability of the transfer in terms of GHS yield and performance. Our initial approach followed a strategy shown in previous studies, in which 2D material devices were integrated onto the CMOS chip's top passivation layer and connected to the underlying circuitry through top-metal pads (**Fig 2A-B**).[26,28,30] However, our results using this approach indicated several obstacles that limited the transfer yield and reliability. One such challenge was the non-planar nature of the CMOS top passivation. Measuring the topography of the chip surface (**Fig 2C**) using optical profilometry (Keyence VR-6000) revealed that the contact pads are recessed by ~1.8 µm relative to the chip surface. Graphene was observed to be prone to tearing at the pad opening edges (**Fig S1A**), consistent with existing reports;[49] furthermore, the steep pad sidewall angle made it challenging to deposit metal contacts to connect the graphene on the top surface to the recessed pads (**Fig S1B-C**). Another obstacle encountered during our top-passivation integration attempts was that the CMOS pads, which are made of Al or Cu in most commercial processes, readily oxidized in ambient conditions and were thus coated by a thin insulating layer of native oxide. This oxide acted as a capacitance in series with the GHS, presenting a high impedance at low frequencies. Test structures shorted by post-fabricated metal (**Fig 2D**) were probed (Cascade PicoProbe) and their reflection coefficient was measured using a vector network analyzer (HP 8753D). Fitting the resulting data to a lumped-element circuit model (**Fig S2**) indicated that the series capacitance associated with each pad was ~10 pF, corresponding to an oxide thickness $t_{ox}$~3 nm consistent with reported values for the self-limiting thickness of native aluminum oxide.[50] This measurement suggested that at sub-GHz frequencies, the series capacitive impedance would dominate the ~kΩ impedance of graphene devices. The rapid re-oxidation kinetics made it challenging to consistently remove the native oxide prior to graphene integration or contact metallization.

To avoid the challenges of integrating graphene onto the top passivation, we investigated the feasibility of integrating graphene onto the inter-layer dielectric (ILD) which separates the layers of the BEOL metal stackup (**Fig 2E**). In this approach, the top metal layer and diffusion barrier are etched, leaving behind a planarized ILD surface with embedded vias defining the device contacts.[51,52] We designed a CMOS chip (**Fig 1c, Fig 2f**) to test the potential advantages of this approach. After removing the top metal and diffusion barrier, we performed atomic force microscopy (AFM) on the ILD surface. The microscopy results (**Fig 2G**) showed that the ILD surface was extremely flat compared to the chip's top passivation, with surface roughness $R_a = 0.75$ nm. In addition to the flatness advantage of ILD integration, we also hypothesized that the tungsten vias would be less prone to oxidation than the top-metal pads, allowing for Ohmic contact between graphene devices and the underlying circuitry. This was tested by selectively shorting sensing sites on the CMOS chip using post-fabricated metal and measuring the resistance at each site using the integrated multiplexing circuitry (**Fig 2h**). The shorted sites had an average resistance of 710 Ω (limited by the on-resistance of the CMOS multiplexing switches), suggesting good contact quality between the vias and the metal shorting bars, while the other sites had an average resistance of 9.4 MΩ. The flatness of the ILD, as well as the capability for Ohmic device contacts using the vias, made this approach appealing for our subsequent GHS-CMOS integration experiments.

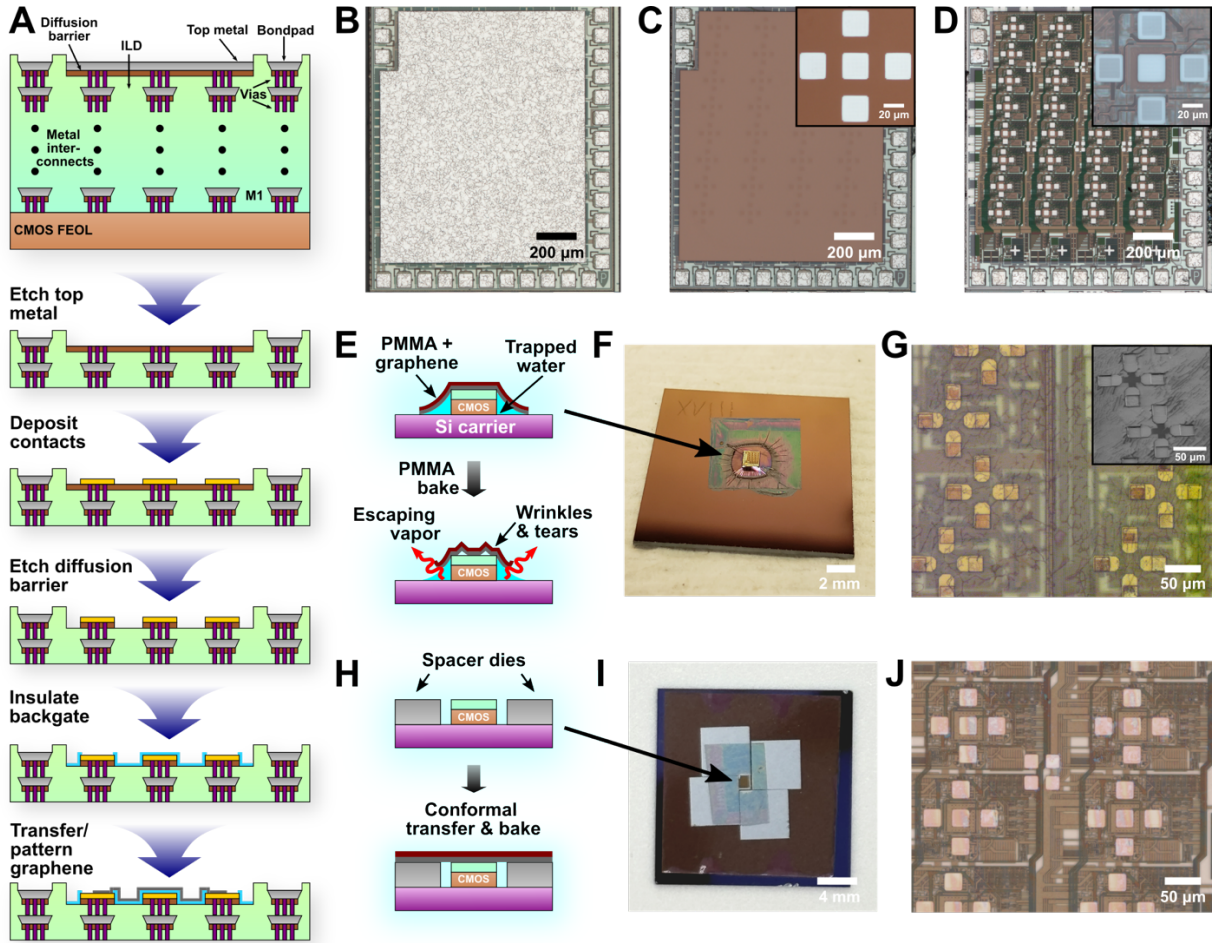

**Figure 3: Graphene-CMOS integration onto the interlayer dielectric (ILD).** A) Schematics of the integration process steps. The topmost schematic shows the complete cross section of the as-received CMOS chip while the subsequent schematics only show the topmost chip layers for compactness. B) Micrograph of the as-received CMOS chip from the foundry. C) Chip micrograph after etching the top metal fill. Inset shows a micrograph of a single sensing site after contact deposition and patterning. D) Chip micrograph after diffusion barrier etching. Inset shows a differential interference contrast micrograph of a single sensing site after backgate insulation. E) Schematic representation of graphene-CMOS integration using the conventional PMMA-assisted wet transfer process. F) Image of the CMOS chip and carrier immediately after graphene transfer using the conventional process. G) Micrograph of sensing sites on the chip after conventional graphene transfer and PMMA removal. The inset shows a scanning electron micrograph of the chip after graphene patterning. H) Schematics of the modified transfer preparation using spacer dies to allow for conformal graphene-substrate contact. I) Image of the CMOS chip and carrier immediately after graphene transfer using the modified preparation. J) Micrograph of several sensing sites on the chip after the modified graphene transfer, prior to PMMA removal.

**Graphene-CMOS integration process**

Graphene Hall sensors were integrated onto the CMOS-ILD using standard semiconductor microfabrication techniques (**Fig 3A**). To process mm-scale CMOS dies using wafer-scale microfabrication instruments, each chip was mounted onto a cm-scale silicon piece using thermal release tape; additionally, all lithography processes were performed using spray coating to avoid edge-bead complications introduced by spin coating onto small dies (**Fig S3**). To maintain CMOS compatibility, the processing temperature of all steps was kept below 250°C to avoid thermally induced dopant migration that could affect circuit performance. This constraint meant that processes such as prolonged plasma etching, high-temperature deposition techniques, and high-temperature annealing steps were avoided.

To integrate GHSs using the ILD approach, the entire sensing region of the chip was designed to be covered by a sacrificial top-metal fill with the chip passivation removed by the foundry (**Fig 3B**). After receiving chips from the foundry, we removed the metal fill using a wet etch process,[53,54] leaving behind a diffusion barrier with exposed vias defining the sensing sites (**Fig 3C, Fig S4**). Metal contacts were then patterned on top of the vias using e-beam evaporation and liftoff (**Fig 3C, inset**). The diffusion barrier was subsequently etched using a peroxide-based wet process to expose the ILD (**Fig 3D**).[55] Metal contact deposition was performed before etching the diffusion barrier, as we found the etchant otherwise corroded the vias (**Fig S5A-D**). The thickness of the metal was chosen to minimize step height while being thick enough to avoid pinhole formation during deposition, thereby avoiding unwanted corrosion (**Fig S5E-I**). To insulate the backgate terminals of the sensors, a thin layer of hafnium dioxide ($HfO_2$) was deposited onto the chips using atomic layer deposition (ALD) (**Fig 3D**, **inset**).[56] The $HfO_2$ covering the other GHS contacts was then selectively removed using photolithography and reactive ion etching (RIE).

A major challenge in the integration process was adapting conventional 2DM wet-transfer methods, involving a poly(methyl methacrylate) (PMMA) backing layer, for compatibility with mm-scale CMOS dies.[57] In the initial transfer attempt, the CMOS chip was mounted onto a silicon carrier substrate (**Fig 3E**). Since the PMMA-graphene membrane was larger than the chip, the membrane draped over the sides, mechanically stressing the graphene around the edges of the chip. Additionally, water from the wet transfer step was trapped underneath the PMMA-graphene membrane; upon heating the substrate to induce PMMA relaxation, the water evaporated, resulting in a pressure buildup that tore the PMMA-graphene membrane (**Fig 3F**). To avoid this complication, the transfer step was optimized by surrounding the CMOS chip by height-matched silicon spacers to ensure that the PMMA-graphene membrane sat flatly on the chip during the transfer process and the subsequent bake step (**Fig 3G**). This preparation minimized wrinkles and strain in the PMMA-graphene membrane; additionally, it provided a path for water underneath the membrane to escape during the bake step, allowing for the PMMA to be heated fully above its glass transition temperature and ensuring conformal contact (**Fig 3H**).

Following the transfer, we used photolithography and oxygen plasma reactive ion etching to pattern the graphene sheet into Hall sensors (W = 10 µm, L = 60 µm) at the pre-defined sites. The chips were then annealed for 1 hour at 225°C under Ar/$H_2$ gas flow to remove surface contaminants.[58] The cleanliness of the transferred chips were inspected using optical microscopy (**Fig 4A**). Dark-field imaging showed minimal indication of processing residues on the chip surface, verifying the cleanliness of the patterning and annealing processes. The sensors were not visible in the optical images due to the low optical contrast of monolayer graphene.[59] To assess the effects of the post-processing steps on sensitive CMOS circuits, the frequency tuning response of an on-chip voltage-controlled relaxation oscillator[60] (**Fig S6**) was measured using a spectrum analyzer (HP 8563E). The tuning response was compared between oscillators on two chips – one measured prior to the integration process and one measured after graphene transfer and patterning (**Fig 4B**). The frequency deviation between the two oscillators was below 3% with an average absolute frequency deviation of 1.19%, suggesting that the integration process had a minimal effect on electronic devices within the CMOS front-end layers.

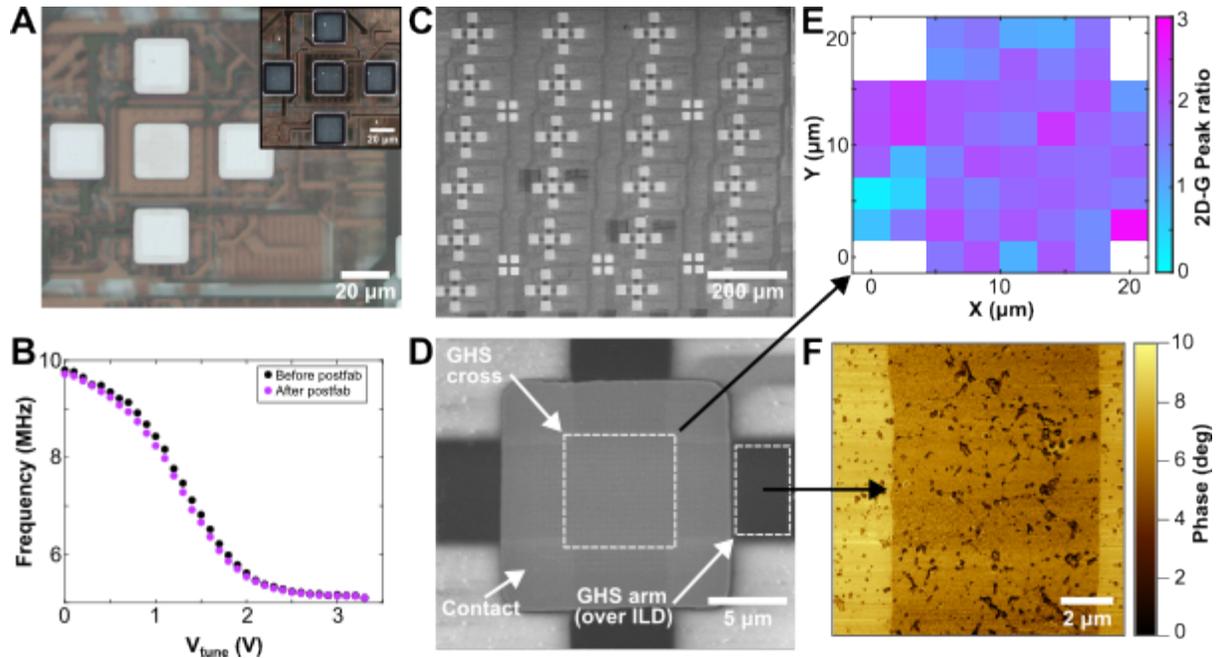

**Figure 4: Characterization of CMOS-integrated graphene Hall sensors.** A) Micrograph of a single sensing site after graphene integration and patterning. Inset shows a darkfield micrograph of the same sensing site. B) Tuning response of an on-chip oscillator circuit on CMOS chips without (black) and with (purple) integrated graphene Hall sensors. C) Scanning electron micrograph of the CMOS-integrated graphene Hall sensor array. D) Magnified scanning electron micrograph of the central cross of a single Hall sensor within the array. E) Raman spectroscopy map of the central cross region showing the ratio between the 2D peak amplitude (located near 2680 cm$^{-1}$) and the G peak amplitude (located near 1580 cm$^{-1}$) following filtering and baseline subtraction. Regions without significant peaks are shaded white. F) AFM phase map of the Hall sensor arm.

The CMOS-integrated graphene Hall sensors were characterized using scanning-electron microscopy (SEM), confocal Raman spectroscopy, and atomic force microscopy (AFM). Environmental-mode SEM revealed dark cross-shaped regions at the sensor sites that corresponded well with the patterning mask (**Fig 4C**). The contrast was higher over the ILD than over the contacts, possibly due to substrate charging (**Fig 4D**).[28] Raman mapping (**Fig 4E**) was used to confirm that the dark regions observed in the SEM images corresponded to monolayer graphene. The raw spectra (**Fig S7**) contained G- and 2D-peaks at 1580 and 2680 cm$^{-1}$, respectively, which are characteristic of graphene.[61] Since the ILD contributes a background signal to the Raman measurement,[28] the spectrum was processed using a custom script to remove the background and de-noise the signal. Following baseline correction, the measured 2D-G peak ratio ($\frac{I_{2D}}{I_G} = 1.65$) was comparable to that of samples transferred onto silicon substrates ($\frac{I_{2D}}{I_G} = 1.54$), suggesting the quality of the graphene was not adversely impacted by the modified transfer process (**Fig S8**).[62] Peaks were only observed in the region of the graphene cross, confirming successful patterning. Furthermore, we used AFM to profile the arms of the GHS cross. The resulting phase images (**Fig 4F**) showed high contrast between the expected graphene regions and the surrounding ILD, providing further evidence for the successful patterning of graphene atop the CMOS chip.

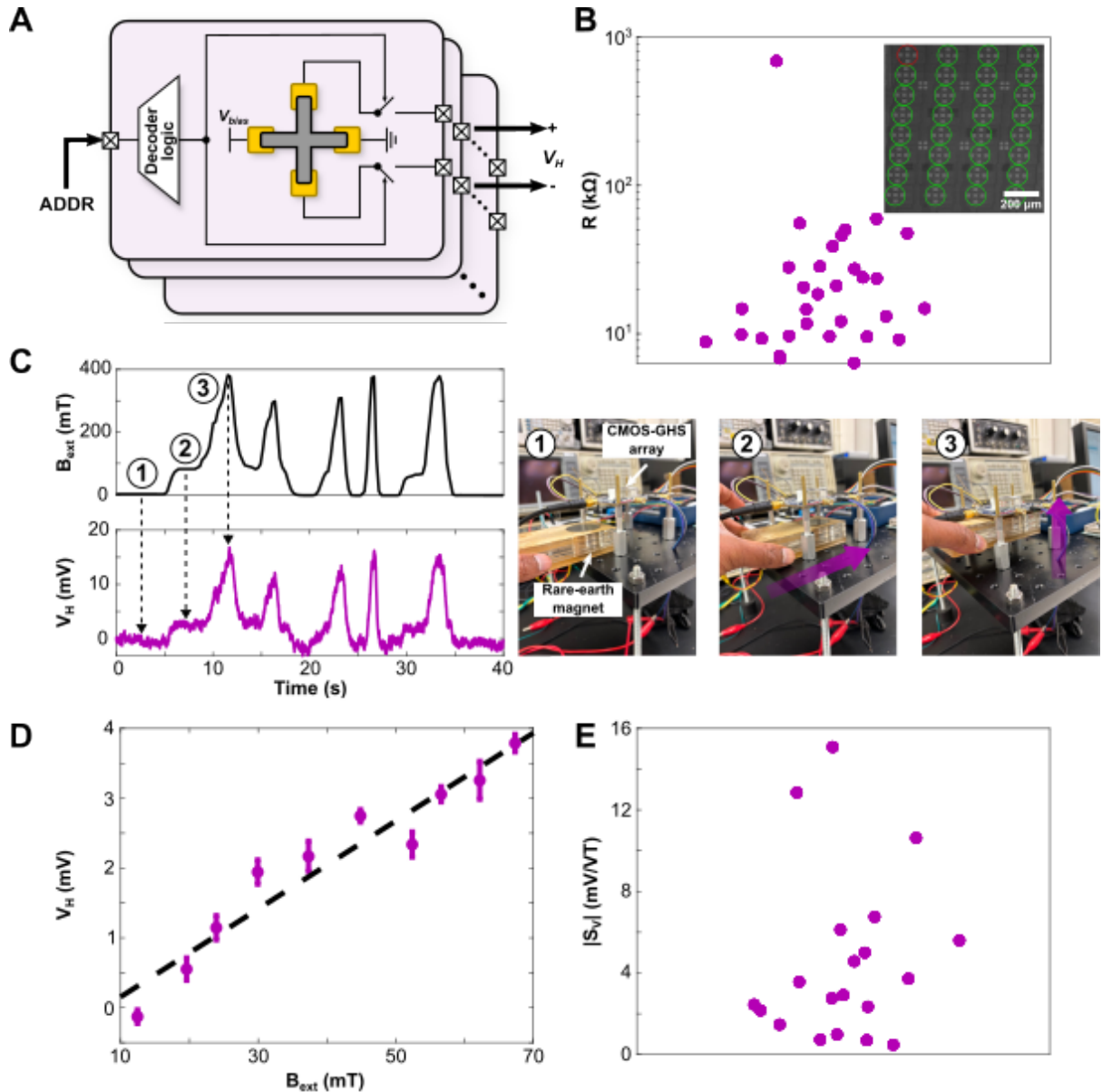

**Figure 5: Electrical & magnetic characterization of CMOS-integrated graphene Hall sensors.** A) Schematic of the CMOS-based multiplexing and readout circuitry used to measure the integrated graphene Hall sensors. B) Scatter plot of measured resistances across the array following GHS integration. The inset shows the locations of intact sensors (R < 100 kΩ) circled in green. C) Measured magnetic field (top left, black) and Hall voltage from a representative sensor (bottom left, purple) induced by a rare-earth magnet (REM) moved underneath the sensing array. The Hall voltage is lowpass filtered (corner frequency 1 Hz) for clarity. The sequence of images (right) shows the positioning of the REM and the CMOS-GHS array at each of the indicated timepoints. D) Dependence of the Hall voltage on magnetic field strength (averaged over 5 measurement repetitions). Error bars indicate the standard error of the mean. E) Scatter plot of Hall sensitivities for responsive devices across the array.

**Magnetic sensing performance of integrated Hall sensors**

After integrating the graphene Hall sensors into the CMOS BEOL, the sensors were electrically measured using the integrated multiplexing circuitry to access each individual sensor (**Fig 5A**). To enhance the array's scalability, address decoding logic was included in each pixel, allowing the input address lines ($N = 5$ bits) to be shared between all ($2^N = 32$) sensing sites. The sensors were biased in the East-West direction using a shared voltage supply generated off chip by an adjustable regulator. The North and South terminals of the selected Hall sensor were connected to two shared output pads using transmission-gate switches, implemented using thick-oxide MOS transistors. To facilitate measurement, the CMOS-GHS chip was integrated with a custom printed circuit board (PCB) including components for power supply regulation, digital bias tuning, and logic level translation (**Fig S9**). The chip was mounted on top of a silicon interposer pre-patterned with Au/Ti fanout traces, which was subsequently mounted onto the PCB. Wirebonds were used to form electrical connections between the chip, silicon interposer, and PCB.

Electrical measurements of the CMOS-integrated GHSs showed evidence of intact graphene devices with high yield (**Fig 5B**). The resistance between the North and South terminals at each site was measured sequentially using a microcontroller (Arduino Due) for sensor selection and a digital multimeter (Agilent 34401A) for readout. 31 out of the 32 measured sites (97%) within the array were found to be intact (defined here as having resistance values below 100 kΩ). The measured yield compares favorably to previous reports of graphene device integrated onto mm-scale CMOS chips, approaching the yield figures reported for cm-scale die and whole wafer integration.[28,31] The intact device resistances ranged from 5.7 kΩ to 59.7 kΩ with a median value of 14.8 kΩ, corresponding to a sheet resistance of 2.47 kΩ/sq (neglecting contact resistance) comparable with existing reports.[63,64]

Magnetotransport measurements of the CMOS-integrated graphene Hall sensors showed that the integrated devices were also magnetically responsive. The Hall response of each sensor was assessed by measuring the voltage difference between the North and South terminals while biasing the East-West direction of the cross with $V_{bias}$ = 3.3 V. A magnetic input signal was provided by rapidly moving a 0.5 T NdFeB rare-earth magnet into and away from the sensing PCB. The magnet diameter (2 in) was chosen to be much larger than the dimensions of the CMOS chip (1.4 mm) to ensure a uniform magnetic field within the sensing area. The transient output voltage from each sensor was directly digitized using a data acquisition system while using a gaussmeter probe to concurrently measure the transient magnetic field. Correlating the two signals showed that the sensor response increased as the magnet was brought closer to the sensing chip (**Fig 5C**); furthermore, the signal reversed polarity when the magnet was flipped, indicating the magnetic nature of the response.

To quantitatively assess the field linearity of the CMOS-integrated GHS response, an electrically tunable external field was supplied by replacing the rare-earth magnet with an electromagnet (Bunting BDE 3020-12). A microcontroller (Arduino Uno) was used to enable/disable the electromagnet using a motor shield driver (Pololu G2). The drive direction was flipped during the measurement to reverse the polarity of the magnetic field and ensure the measured response was magnetic in nature. As the electromagnet was turned on and off, the sensor output voltage was measured directly using a multimeter to quantify the offset voltage and Hall response. The field linearity was measured by adjusting the driver supply voltage between 8 and 24 V, changing the magnetic field strength up to ~70 mT. Representative results for a single sensor (**Fig 5D**) show that as the magnetic field strength was increased, the Hall voltage increased linearly ($R^2 = 0.93$) with a voltage-normalized sensitivity of 19 mV V$^{-1}$ T$^{-1}$.

Conducting magnetic measurements across the CMOS-integrated GHS array provided insight into device yield and variability. The correlation between the measured transient magnetic field

(applied using the rare-earth magnet) and the Hall signal was calculated for each sensor's response and a threshold was applied to differentiate magnetically responsive devices from non-responders. Using this approach, 20 sensors (63%) within the array were found to be responsive (**Fig 5E**). This figure exceeds the functional device yield reported in previous die-level integration studies (42% and 52%, respectively).[28,31] The number of magnetically responsive devices was found to be lower than the number of intact devices. The discrepancy is hypothesized to have occurred because the Hall measurement required four intact device terminals, whereas the resistance measurement only required two terminals (**Fig S10**). For each responsive graphene Hall sensor, the voltage sensitivity $S_V = V_H/(B_{ext}V_{bias})$ was estimated, where $V_H$ is the measured Hall voltage and $B_{ext}$ the measured external field from the REM. The voltage sensitivity of the responsive devices exhibited significant device-to-device variability, ranging from 0.48 mV V$^{-1}$ T$^{-1}$ to 15.1 mV V$^{-1}$ T$^{-1}$ with a median value of 3.2 mV V$^{-1}$ T$^{-1}$. Heterogeneity was also observed in the offset voltages of the responsive devices (**Fig S11**), ranging from -1.41 V to 0.79 V with a median absolute offset of 148 mV. The variability observed across the array was somewhat expected given that the GHS response is known to be sensitive to doping variations induced by local non-idealities such as interactions with charge traps in the substrate, adsorbed dopants, grain boundaries, mechanical strain, lithographic imperfections, and patterning residues.[47,62,65] Further evidence for doping inhomogeneity across the chip was suggested by the observation that devices on the chip had differing Hall response polarities, despite the field and bias polarities being identical for both devices (**Fig S12**). This behavior can occur between graphene Hall sensors that are doped on opposite sides of the Dirac point, leading to the Hall effect being mediated by different charge carrier polarities in each device.[66]

## CONCLUSION

This work shows for the first time that graphene Hall sensors can be monolithically integrated with commercial silicon CMOS, enabling dense arrays that stand to improve state-of-the-art performance in several domains of magnetic sensing. These include expanding the scope of magnetic profiling techniques such as scanning Hall probe microscopy (SHPM)[41,67] and mapping electric current distributions[43] by breaking the tradeoff between field of view and spatial resolution. The Hall elements used in these applications are typically small (1 μm x 1 μm) to analyze microscale features, resulting in lengthy scan times (~4 hours for a 1 mm$^2$ sample);[41] by contrast, a GHS array with $N$ = 20 elements (as shown in this work) could scan the same area in 12 minutes. Additionally, CMOS-integrated GHS arrays are an ideal sensing platform for parallelized magnetic flow cytometry with high sensitivity, enabling the analysis of high-volume samples in biosensing scenarios such as the detection of circulating tumor cells (CTC) for cancer management[44,46] and the detection of pathogenic bacteria for environmental and food monitoring.[68,69,70] Since the targets in these applications can be sparse (<100 targets per mL of sample), large sample volumes (>10 mL) are required to avoid subsampling error; however, the flow rate using a single sensor must be slow enough to allow sufficient signal integration.[70] Using a CMOS-integrated GHS array with 20 elements, the measurement time for a 10 mL sample can be reduced from 10 hours to 30 minutes, making it feasible to perform the assay in the lab or in the field. The measurement times in these applications can be further reduced by incorporating more sensors into the array by increasing the chip area or by decreasing the sensor-to-sensor pitch.

While this work demonstrates the advantages of GHS-CMOS integration in terms of array density, monolithic integration also offers the opportunity to integrate graphene Hall sensors with CMOS-based control and readout circuitry. The results from this work and in other scaled 2DM-CMOS systems suggest that device-to-device variability places a significant limit on overall array performance. Implementing per-pixel GHS response tuning using CMOS allows for heterogeneity within the array to be compensated, improving sensing uniformity;[62] furthermore, dynamic or adaptive modulation of each GHS's response can be used to perform in-sensor computation and signal processing, enhancing the functionality of the array.[71,72] Furthermore, compact CMOS readout circuits can be used to mitigate undesirable features of the GHS output signal such as low-frequency noise and zero-field offset voltages, improving their sensing performance.[73,74]

Although 2DM-CMOS integration is explored here within the context of GHS arrays, the processing strategies discussed here are broadly applicable to devices made from other 2D materials and serving other applications. Monolithic integration of 2D materials with commercial silicon CMOS has been relatively underexplored, despite the clear benefits in terms of device density and functionality, due to reliability concerns about 2DM transfer onto CMOS dies designed in multi-project wafer runs; on the other hand, relatively few research groups have the resources required to pursue wafer-scale integration. Exploratory research efforts will be critical in realizing the full potential of 2DM-CMOS integrated systems by developing methods to maximize integration yield, introducing 2D materials beyond graphene into the CMOS BEOL, and increasing integration density by miniaturizing CMOS-integrated 2DM devices to the submicron scale. These device-level innovations will need to be complemented by research from chip designers into developing new CMOS circuit architectures to mitigate the non-idealities of 2DM devices while harnessing their performance advantages and new capabilities. By studying the challenges of die-scale integration and introducing possible solutions to improve integration reliability, this work may help to democratize research into 2DM-CMOS integration by providing a path towards widespread investigations using readily available CMOS integrated circuits.

## RESOURCE AVAILABILITY

The data and analysis code that support the findings of this study are available from the corresponding authors upon reasonable request.

## ACKNOWLEDGEMENTS

This work was supported by the National Cancer Institute (R21CA236653, R33CA278551, CA261608), the National Institute of Mental Health (R33-NIMH-118170), and the National Institute of Allergy and Infectious Diseases (R33-AI147406). This work was carried out in part at the Singh Center for Nanotechnology, which is supported by the NSF National Nanotechnology Coordinated Infrastructure Program under grant NNCI-2025608. We thank Jaeung Ko, Yeonjoon Suh, Chengyu Wen, Qicheng Zhang, and staff members at the Singh Center for Nanotechnology for their helpful suggestions regarding device fabrication and graphene characterization. The authors thank Muse Semiconductor and members of the EPM Laboratory at the University of Pennsylvania for helpful discussions regarding CMOS chip design and measurement. We also thank Neha Srikumar for providing constructive feedback on the manuscript.

## AUTHOR CONTRIBUTIONS

V.I., D.I., and F.A. conceptualized the study. V.I. performed all experiments and wrote the manuscript. N.S. helped with the development of fabrication procedures. A.T.C.J. provided resources for graphene synthesis and transfer. All authors contributed to the discussion of the results and reviewing the manuscript.

## DECLARATION OF INTERESTS

V.I., D.I., and F.A. are listed as co-inventors on a provisional patent filed with the US Patent and Trademark Office (application number 63/673,547; filing date 19 July 2024) covering aspects of the proposed 2DM-CMOS integration strategy.

**METHODS**

**CMOS chip design**

The CMOS integrated circuits discussed in this work were implemented in a Taiwan Semiconductor Manufacturing Company (TSMC) 180nm bulk CMOS process. Schematic-level design and chip layout were performed using Cadence Virtuoso (Cadence Design Systems). Calibre (Siemens) was used for design rule checks, layout verification, and parasitic estimation. The chips were fabricated by TSMC using a multi-project wafer shuttle program (Muse Semiconductor).

**Graphene-CMOS integration process**

All post-fabrication steps were performed using equipment at the Quattrone Nanofabrication Facility within the Singh Center for Nanotechnology at the University of Pennsylvania. UV lithography was used in all steps to define processing regions on the chip. Photomask patterns for lithography were designed in KLayout using layout GDS files exported from Cadence as a reference. A Heidelberg DWL66+ laser writer with a 2 mm write head was used to write the patterns into a 5'' glass photomask coated with chromium and AZ1500 photoresist. The mask was developed in AZ300 MIF developer (Microchemicals) for 90 s and the exposed chromium was etched for 2 minutes in Chromium Etchant 1020AC (Transene). Remaining photoresist was stripped from the mask by sonication in Remover 1165 (Microposit) heated to 60°C, followed by rinses in acetone and isopropyl alcohol (IPA).

Each CMOS die was attached to a 1 cm x 1 cm silicon carrier chip (University Wafer) using thermal release tape (Nitto Denko Revalpha 31950E) and rinsed with acetone and IPA prior to post-fabrication. To remove the top-metal sacrificial layer, the chip was first coated with 4 µm of positive photoresist (Shipley S1800) deposited using a spray coater (Suss MicroTec AS8 AltaSpray) and baked for 3 minutes at 95°C. The photoresist above the sensing region was exposed with 158

mJ/cm$^2$ exposure energy using an ABM 3000HR mask aligner in proximity mode and developed using AZ300 MIF for 1 minute, leaving the bondpads covered with photoresist. The top-metal sacrificial layer was then etched with Transene Aluminum Etch Type A heated to 50°C for 25 minutes followed by a water rinse for 1 minute. The remaining photoresist was removed by rinses in Remover 1165, acetone, and IPA.

For metal contact formation, 3 µm of negative photoresist (APOL-LO 3200) was deposited onto the chip using a spray coater and baked for 1 minute at 115°C. The chip was exposed using a Suss MicroTec MA6 Gen3 mask aligner in hard contact mode with 270 mJ/cm$^2$ exposure energy, followed by a 2-minute post-exposure bake at 115°C and development with AZ300 MIF for 45 seconds. Prior to metal deposition, a descum step in oxygen plasma (50 W RF, 100 sccm $O_2$, 30 s) was performed using a parallel-plate reactive ion etcher (Jupiter March II RIE) to remove photoresist residues. A metal stack of 10 nm Ti and 50 nm Pd was then deposited by e-beam evaporation (Kurt Lesker PVD75) at a rate of 2 Å/s each. Liftoff was performed overnight in Remover 1165 heated to 60ºC without sonication to preserve smooth metal edges, followed by rinses in acetone and IPA. The exposed diffusion barrier regions were then etched using an ammonia-peroxide mixture (1:2:5 28% ammonium hydroxide: 30% hydrogen peroxide: deionized water) for 30 minutes at room temperature followed by a water rinse for 1 minute.

The backgate terminal was insulated by depositing hafnium oxide onto the chips using an atomic layer deposition (ALD) process (Cambridge NanoTech) with the chamber temperature set to 150°C. Each deposition cycle consisted of alternating pulses of HFDMA (0.4 s) and $H_2O$ (0.015 s) with 8 s clearing time between pulses. Deposition thickness was characterized with a Filmetrics F40 instrument on test chips included within the ALD chamber, since the complex BEOL dielectric stack made it challenging to reliably measure $HfO_2$ film thickness on the CMOS chips. Etching windows over the GHS contacts were defined using 4 µm S1800 photoresist deposited by spray coating, exposed with UV photolithography (Suss MicroTec MA6 Gen3) at 120 mJ/cm$^2$, and developed with AZ300 MIF for 1 minute. The Hall sensor contacts were exposed by etching $HfO_2$ using RIE (Oxford 80+) for 5 minutes at 200 W with 40 sccm $CHF_3$ and 20 sccm Ar. Photoresist hardened by the RIE step was removed with an oxygen plasma RIE (150 W RF, 100 sccm $O_2$, 3 mins). Remover 1165, acetone, and IPA were used to remove remaining resist residues.

Prior to graphene transfer, the CMOS chip was attached to a 2 cm x 2 cm silicon carrier chip using thermal release tape (Nitto Denko Revalpha 31950E) and surrounded with four 6 mm x 6 mm silicon dies each with a thickness of 250 µm (University Wafer). Monolayer graphene was grown onto 99.9999% copper foil (Alfa Aesar) in a chemical vapor deposition system for 20 minutes at 1020ºC under 10 sccm $CH_4$ + 80 sccm $H_2$ flow. The foil was coated with 300 nm PMMA A4 950 and cut into pieces for transfer. The uncoated side of the foil was etched with oxygen plasma (50 W RF, 100 sccm $O_2$, 15 s) using RIE (Jupiter March II) to remove graphene on the back side. The copper foil was then etched for 20 minutes using ferric chloride (Transene) to release the graphene-PMMA membrane. The membrane was drawn from the ferric chloride solution using a plastic film, rinsed 3x in DI $H_2O$ and landed onto the target CMOS chip. The transferred chip was dried in air for 2 hours then baked at 150°C for 2 hours to heat the PMMA past its glass transition temperature. The chip was then soaked overnight in acetone to remove PMMA. Hall sensors were defined using 2 µm S1800 photoresist deposited by spray coating, exposed with UV photolithography (Suss MicroTec MA6 Gen3) in hard contact mode at 80 mJ/cm$^2$, and developed with AZ300 MIF for 45 seconds. Exposed graphene was patterned using an oxygen plasma RIE (50 W RF, 100 sccm $O_2$, 30 s) and the patterning photoresist was removed with Remover 1165, acetone, and IPA. The devices were annealed for 1 hour at 225ºC under 1000 sccm Ar + 250 sccm $H_2$ flow to improve contact resistance and remove patterning residues.

Processed chips were inspected by environmental scanning electron microscopy (E-SEM) using a Quanta 200 SEM operated in low-vacuum mode with 5 kV beam voltage. Atomic force microscopy was performed using a Bruker Icon instrument operating in tapping mode. Raman spectra were obtained using an Ntegra NT-MDT upright confocal Raman scanning microscope with a 532nm green excitation laser and an excitation power of 0.75 mW. Raman spectra were processed using a custom MATLAB script to remove background and de-noise the signal. The baseline was first estimated by passing the Raman spectrum through a median filter (150 sample points); the baseline was then subtracted from the raw spectrum. The baseline-corrected spectrum was filtered using a 3$^{rd}$ order Savitzky-Golay smoothing filter. Peak-finding was then applied to estimate the intensity, location, and width of the D, G, and 2D peaks. Finally, the signal was normalized to the amplitude of the highest peak.

**Interposer fabrication**

Interposers were fabricated on 4'' silicon wafers pre-coated with 275 nm wet thermal oxide (University Wafer). Liftoff regions were defined using a bilayer photoresist stack (LOR3A/S1813) and UV exposure at 120 mJ/cm$^2$ (Suss MicroTec MA6 Gen3). Prior to metal deposition, a descum step in oxygen plasma (50 W RF, 100 sccm O$_2$, 30 s) was performed using a parallel-plate reactive ion etcher (Jupiter March II RIE) to remove photoresist residues. A metal stack of 10 nm Ti and 100 nm Au was then deposited by e-beam evaporation (Kurt Lesker PVD75) at a rate of 2 Å/s each. Liftoff was performed for 30 minutes in Remover 1165 heated to 60ºC using sonication, followed by rinses in acetone and IPA. The wafers were then diced into 2 cm x 2 cm interposer chips using a dicing saw (Advanced Dicing Technologies 710S) with a nickel-hub blade.

**Setup for electrical and magnetic measurements**

CMOS chips were mechanically bonded to the interposers using crystal bond (Ted Pella 555) and connections were made to the fanout traces using gold wirebonds. The interposers were then wire-bonded to a custom printed circuit board (PCB) for measurement. The PCB was designed using Altium Designer (Altium Ltd.) and fabricated by PCBWay. The chip and wirebonds were covered by a plastic Petri dish (CellTreat Scientific) to avoid dust adsorption onto the chip and to protect the chip from mechanical damage during measurement. The PCB was screwed into an optical breadboard (ThorLabs) using non-magnetic standoffs (McMaster-Carr) for mechanical stability. The breadboard was suspended above the measurement table with rubber-footed standoffs to avoid vibrations. When not being measured, chips were kept in a desiccator to avoid exposure to ambient water.

**Hall sensitivity measurements**

The Hall sensitivity of all elements within the array were quantified in a single transient measurement using a semi-automated procedure. An Arduino Due was used to control the sensor bias voltage through a digital-to-analog converter (DAC) onboard the PCB; the Arduino also incremented the selected sensor address upon prompting by the user. The magnetic input signal was provided using a NdFeB rare-earth disk magnet (K&J Magnetics DY0X0-N52) with a 2 in diameter and 1 in axial height. To simplify handling, the magnet was encased in a laser-cut acrylic housing with one face of the magnet exposed.

After each sensor was selected using the Arduino, the disk magnet was rapidly inserted and removed underneath the sensing PCB; the procedure was then repeated with the field polarity reversed by physically flipping the magnet. The transient output voltage from the chip was digitized using a data acquisition (DAQ) system (National Instruments NI-6361) at a sampling rate of 1 kHz. The chip output voltage was directly connected to the DAQ without using amplification to capture the zero-field offset voltage and the Hall response of each sensor. A gaussmeter probe (AlphaLab GM2) was taped to the backside of the PCB directly underneath the chip to independently measure the magnetic field at a sampling rate of 4 Hz.

After data collection, the transient signals were processed by a custom MATLAB script to identify responsive devices and quantify Hall sensitivity. The chip output voltage and magnetic field transient signals were aligned and segmented, yielding transient signals for each sensor. The output voltage was low-pass filtered (corner frequency 2 Hz) and its initial voltage offset was subtracted to quantify the Hall voltage, while the magnetic field signal was interpolated to match the sampling rate of the Hall voltage. The cross-correlation between the Hall voltage and the magnetic field was then calculated, with responsive sensors being identified as those with correlation peaks at zero lag between the signals. The Hall sensitivity was quantified for each responsive sensor by calculating the difference between each sensor's output at maximum and minimum magnetic field (averaged over 100 ms) and normalizing by the change in magnetic field.

# Supporting information

# High-density and scalable graphene Hall sensor arrays through monolithic CMOS integration

Vasant Iyer[1,2*†], Nishal Shah[3], A.T. Charlie Johnson[4], David A. Issadore[1,3*], Firooz Aflatouni[1*]

[1] Department of Electrical and Systems Engineering, School of Engineering and Applied Science, University of Pennsylvania, Philadelphia, Pennsylvania, United States

[2] Querrey Simpson Institute for Bioelectronics, Northwestern University, Evanston, Illinois, United States

[3] Department of Bioengineering, School of Engineering and Applied Science, University of Pennsylvania, Philadelphia, Pennsylvania, United States

[4] Department of Physics and Astronomy, School of Arts and Sciences, University of Pennsylvania, Philadelphia, Pennsylvania, United States

*Co-corresponding author: vasant@northwestern.edu

*Co-corresponding author: issadore@seas.upenn.edu

*Co-corresponding author: firooz@seas.upenn.edu

[†]Lead contact

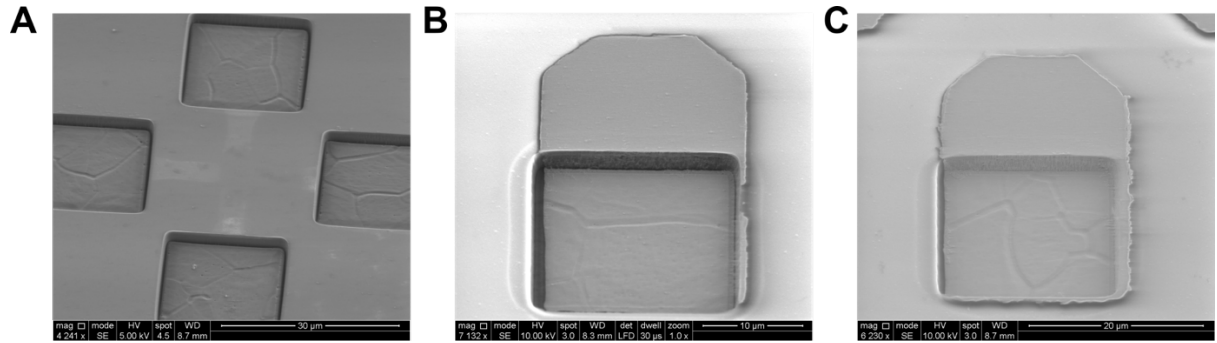

**Figure S1: Effect of top passivation topography on graphene integration and contact metallization.** A) Scanning electron micrograph of a graphene cross on the top surface of a CMOS chip surrounded by top metal pads. B) Scanning electron micrograph of a CMOS top-metal pad with contact metallization (10 nm Ti + 100 nm Au) added using e-beam evaporation (Kurt Lesker PVD75) and liftoff. The anisotropic metal deposition results in smooth contact edges after liftoff but limited sidewall coverage, breaking the contact between the pad surface and the passivation surface. C) Scanning electron micrograph of a CMOS top-metal pad with contact metallization (50 nm Ti + 300 nm Au) added using e-beam evaporation and liftoff. The thicker deposition results in complete sidewall coverage but poor liftoff, resulting in jagged contact edges.

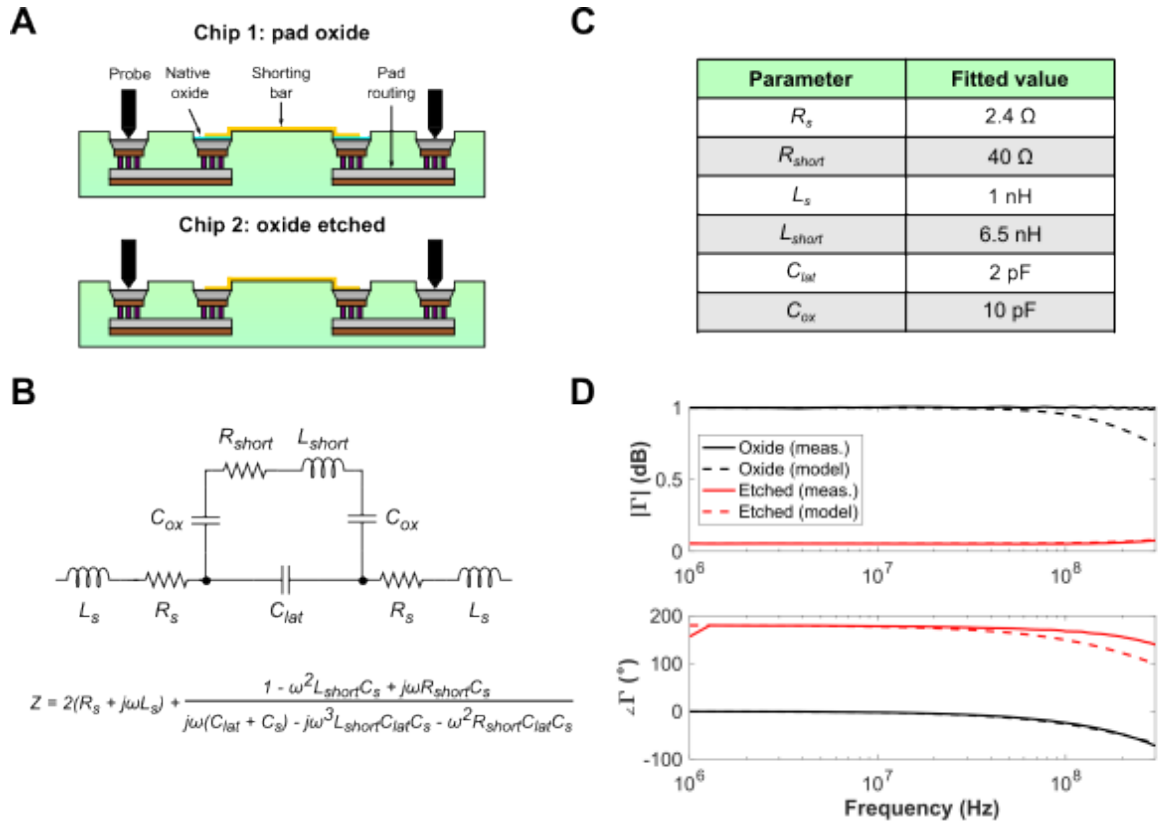

**Figure S2: Quantification of CMOS top-metal pad capacitance due to oxidation.** A) Schematic of the two CMOS chips used to measure the pad oxide capacitance. In both chips, a shorting bar made of evaporated Au/Ti was added between two contact pads (each 20 μm x 20 μm), each of which was connected using lower-layer routing to probing pads. In the control chip, the oxide was etched within the metal deposition tool after pump-down using an argon plasma etch (20 minutes, 4 sccm Ar, 100 W RF). B) Equivalent circuit model and impedance equation of the setup shown in Fig S2A. In the equation, $C_S = C_{ox}/2$ represents the series combination of the oxide capacitances of both contact pads. C) Values used to fit the measured impedance data to the model shown in Fig S2B. The values were chosen to simultaneously fit the measured results from both chips, using $C_{ox} = 0$ on Chip 2. D) Measured and fitted reflection coefficient measurements of the shorted test structures. The model accuracy drops above 100 MHz, likely due to additional parasitics introduced by the long measurement cables.

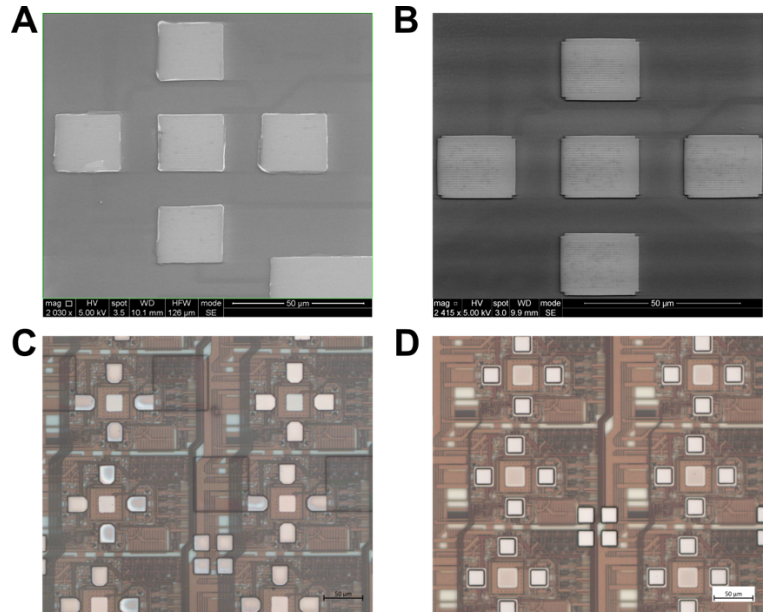

**Figure S3: Optimized UV lithography conditions for CMOS die post-processing.** A) Scanning electron micrograph of metal contacts (10 nm Ti + 25 nm Pd) formed by liftoff using positive photoresist (Shipley S1800 series, thickness 4 µm). The positive photoresist exposure profile made it challenging to perform liftoff without using ultrasonication, resulting in jagged contact edges. B) Scanning electron micrograph image of metal contacts (10 nm Ti + 25 nm Pd) formed by liftoff using negative photoresist (APOL-LO 32, thickness 3 µm). The negative resist undercut profile allowed for liftoff without requiring agitation or ultrasonication, resulting in smooth contact edges. C) Micrograph of lithography results using spin-coated photoresist (Shipley S1813, target thickness 2.5 µm). Due to the small CMOS die size, the photoresist did not spin uniformly, resulting in significant variations in required exposure energy across the chip. D) Micrograph of lithography results using spray-coated photoresist (Shipley S1800 series, thickness 4 µm). Spray coating allowed for uniform photoresist coverage and consistent exposure requirements across the chip.

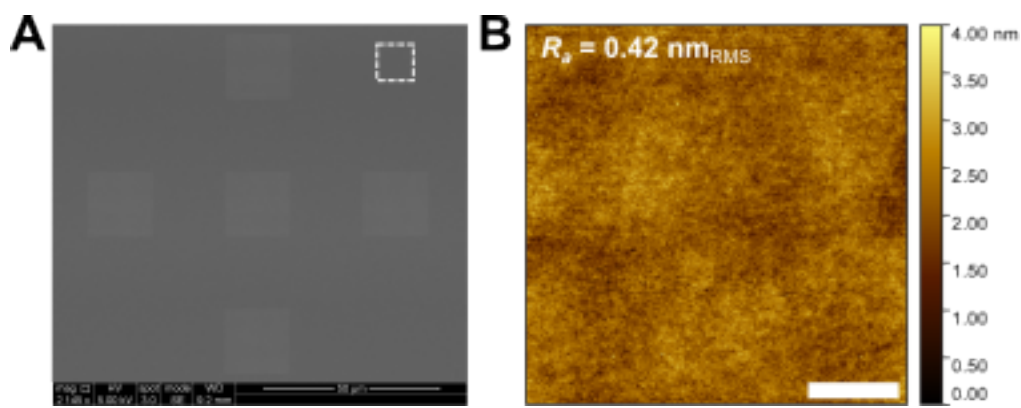

**Figure S4: Characterization of CMOS diffusion barrier.** A) Scanning electron micrograph of a single sensing site following top-metal etching. The dashed box indicates the location of the atomic force microscopy scan shown in Fig S4B. B) Atomic force micrograph of the diffusion barrier surface shown in Fig S4A (scale bar 2 µm).

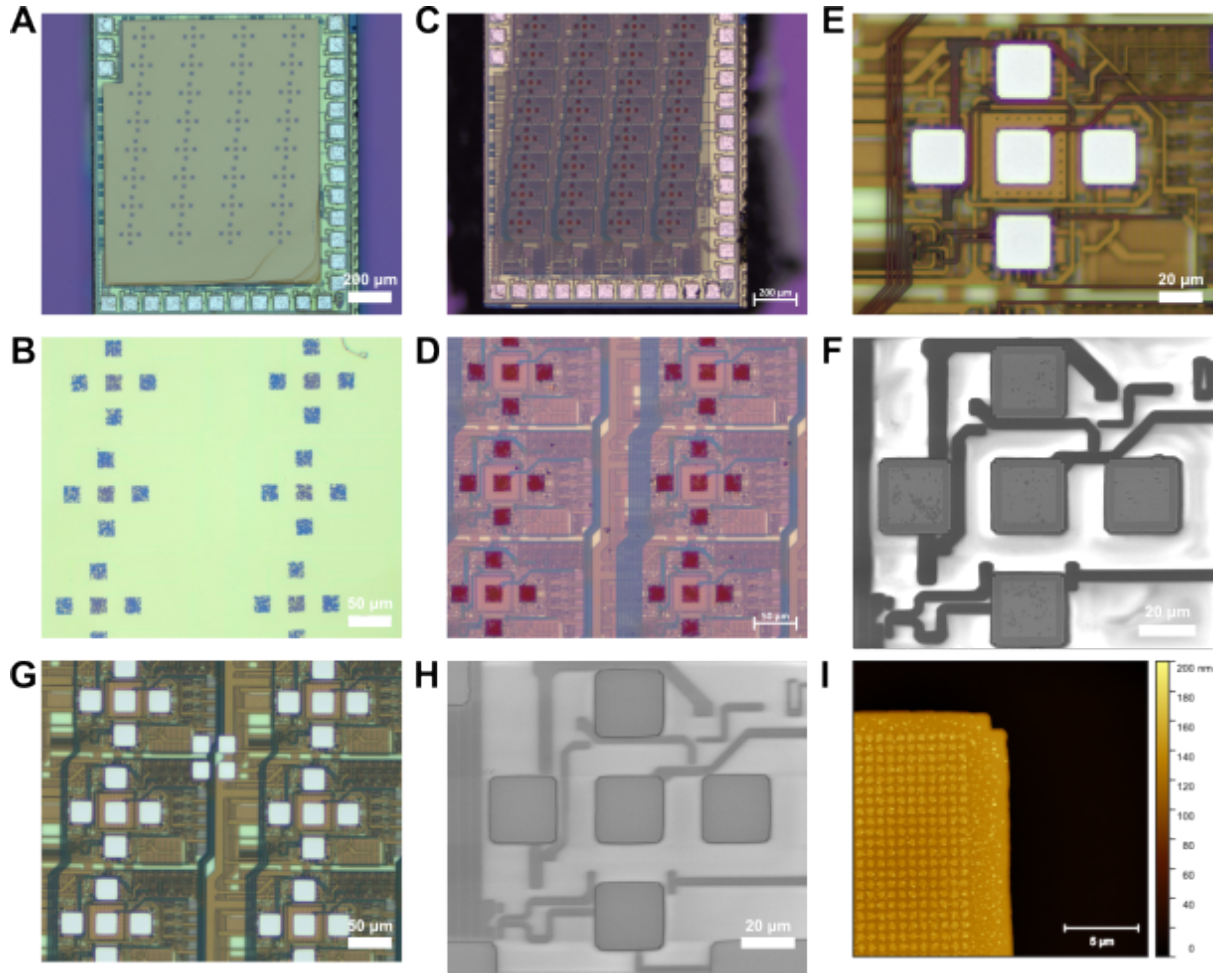

**Figure S5: CMOS diffusion barrier etch process development.** A) Micrograph of the CMOS chip after etching the diffusion barrier for 15 minutes. B) Zoomed-in micrograph of the chip shown in Fig S5A showing partial barrier etch and via attack. C) Micrograph of the CMOS chip after etching the diffusion barrier for 25 minutes. D) Zoomed-in micrograph of the chip shown in Fig S5C showing more complete diffusion barrier etch and via degradation. E) Micrograph of a single sensing site in which metal contacts (10 nm Ti + 25 nm Pd) were deposited on the vias prior to diffusion barrier etching for 25 minutes. However, the vias still show evidence of attack. F) Scanning electron micrograph of the sensing site shown in Fig S5E. The darkened clusters of vias suggest that the etchant accesses the vias through pinholes in the metal contact. G) Micrograph of sensing sites protected by thicker metal contacts (10 nm Ti + 50 nm Pd) prior to diffusion barrier etching for 30 minutes. H) Scanning electron micrograph of a single sensing site protected by thicker metal contacts, showing complete barrier etching and no evidence of via attack. I) Atomic force micrograph of a metal contact and the surrounding inter-layer dielectric following successful diffusion barrier etching.

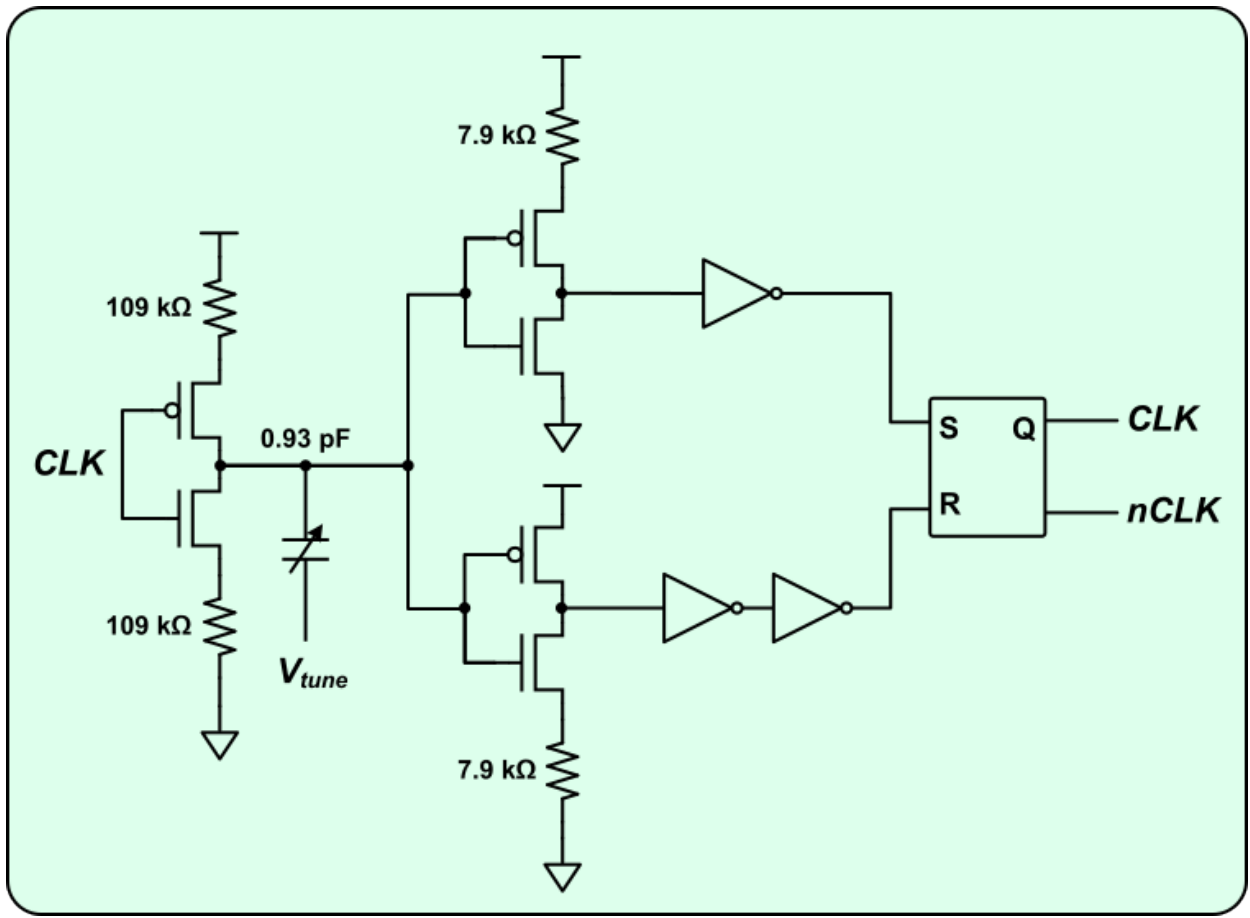

**Figure S6: Schematic of the voltage-controlled relaxation oscillator used to assess CMOS circuit performance after graphene integration.**

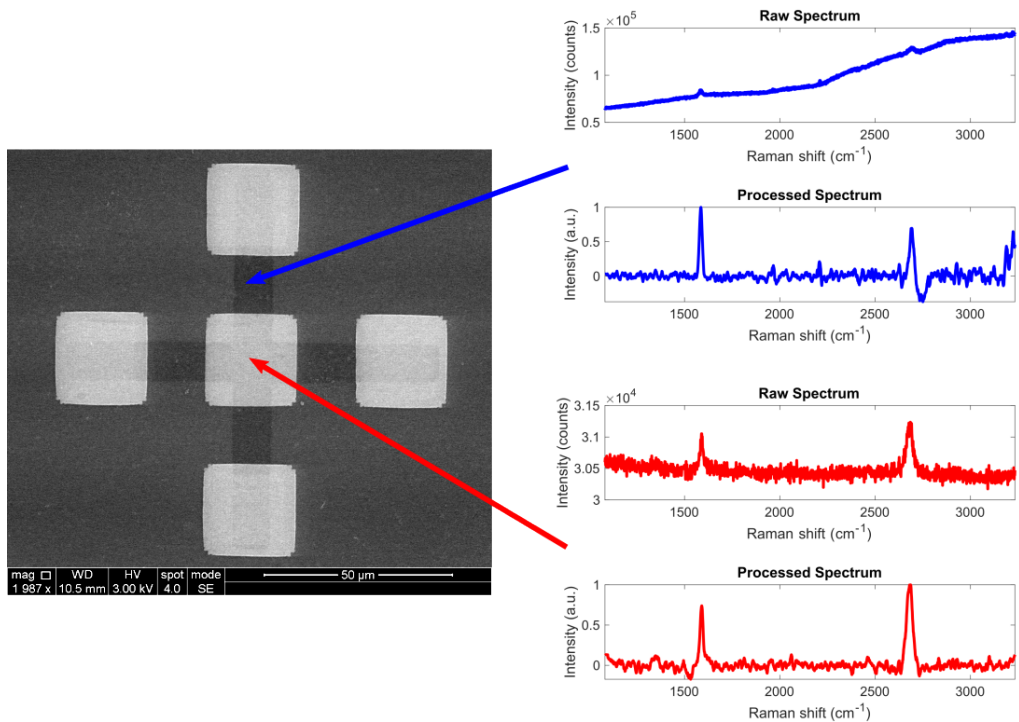

**Figure S7: Raman spectroscopy of CMOS-integrated graphene Hall sensors.** Spectra taken on portions of the graphene Hall sensor lying directly on the inter-layer dielectric (blue) showed a significant baseline signal associated with the dielectric, making it challenging to estimate 2D-G peak ratio in these regions. By contrast, the spectra taken over the metal contacts (red) showed substantially lower background signal, allowing quantification of the 2D-G peak ratio.

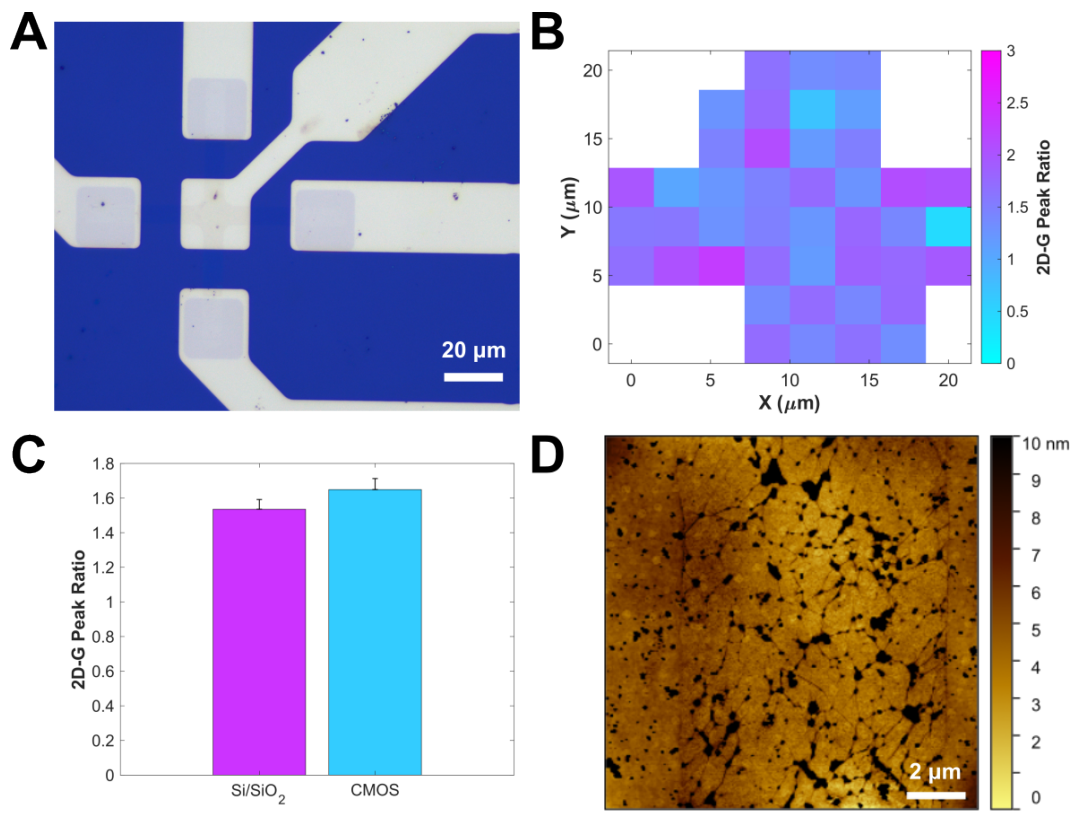

**Figure S8: Additional Raman and atomic force microscopy characterization of CMOS-integrated graphene Hall sensors.** A) Micrograph of a reference graphene Hall sensor integrated onto a conventional SiO$_2$/Si substrate. B) Raman spectroscopy map of the reference graphene Hall sensor showing the 2D-G peak ratio in the central cross region. C) Comparison of average 2D-G peak ratio between graphene Hall sensors integrated on SiO$_2$/Si (pink) and CMOS (blue). Error bars indicate standard error of the mean across each scan. D) Atomic force microscopy height plot of the Hall sensor arm.

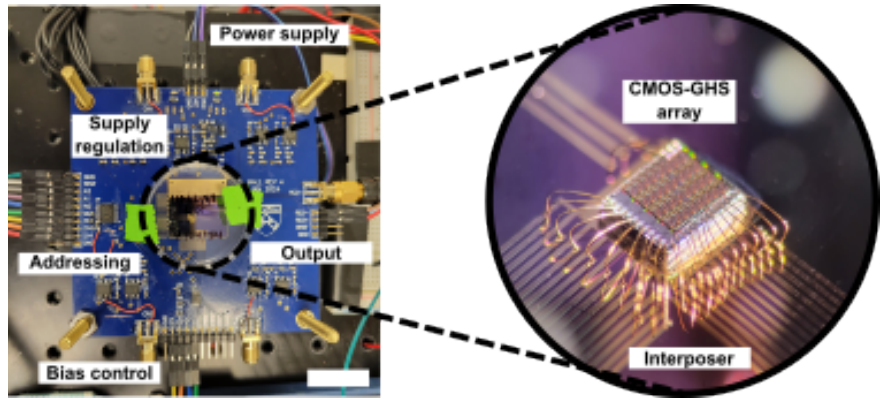

**Figure S9: Setup for electrical and magnetic measurements of the CMOS-GHS array.** Image of the printed circuit board with key functional blocks labeled. Scale bar 2 cm. Inset shows the CMOS-GHS array chip wirebonded to fanout traces on a silicon interposer.

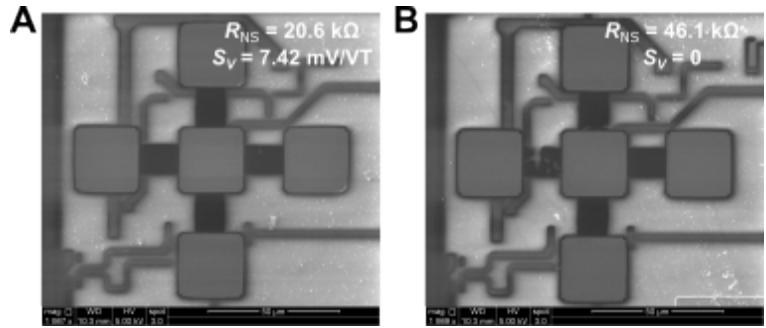

**Figure S10: Imaging analysis of non-responsive graphene Hall sensors.** A) Scanning electron micrograph of an intact, responsive graphene Hall sensor. B) Scanning electron micrograph of a non-responsive graphene Hall sensor with a finite resistance value in the North-South direction. The lack of response can be explained by visible graphene tears in the West-East direction, leading to improper device biasing.

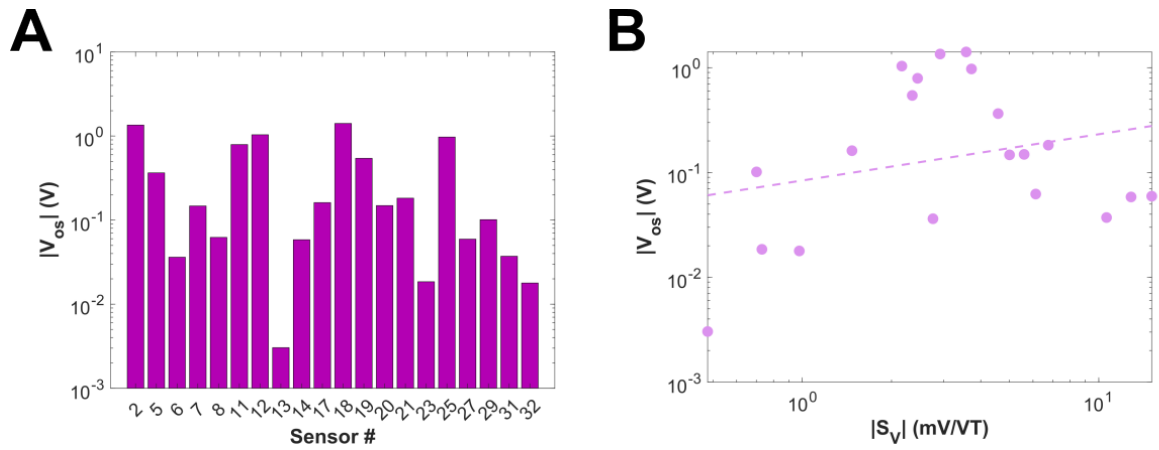

**Figure S11: Offset voltage characterization of CMOS-integrated graphene Hall sensors.** A) Measured offset voltages of responsive Hall sensors on a single CMOS chip. B) Correspondence between the measured Hall sensitivity and offset voltage of responsive Hall sensors. The trendline indicates a weakly positive correlation ($\rho$ = 0.26).

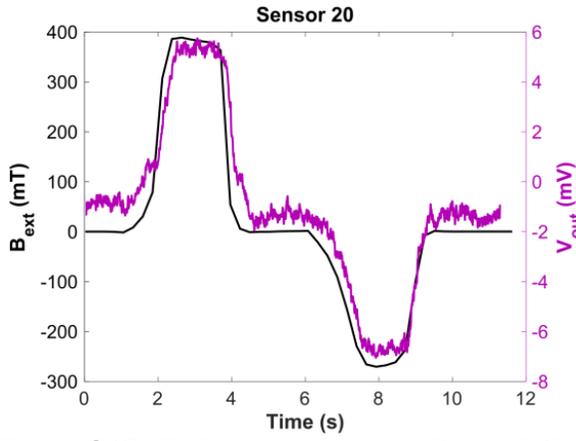 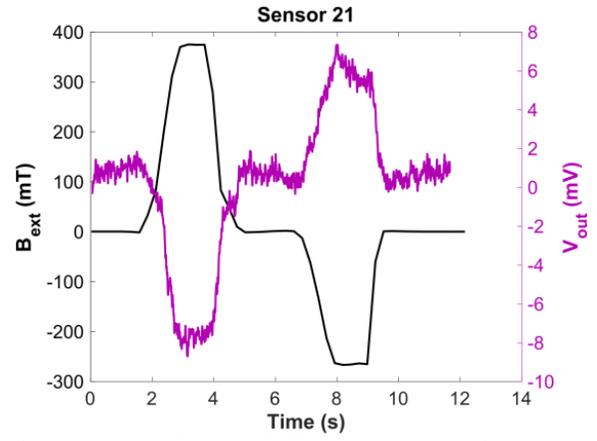

**Figure S12: Hall response polarity variation between graphene Hall sensors on the same chip.**